\documentclass[prl,showpacs,reprint]{revtex4-1}

\usepackage{epsfig,amsmath,amssymb}
\usepackage[outdir=./]{epstopdf}

\def\fig#1{Fig.~\ref{#1}}
\def\eq#1{Eq.~(\ref{#1})}

\begin{document}
	
\title{Gravitational Casimir effect} 

\author{James Q. Quach} 
\email{quach.james@gmail.com}
\affiliation{Institute for Solid State Physics, University of Tokyo, Kashiwa, Chiba 277-8581, Japan}

\begin{abstract}
We derive the gravitonic Casimir effect with non-idealised boundary conditions. This allows the quantification of the gravitonic contribution to the Casimir effect from real bodies. We quantify the meagreness of the gravitonic Casimir effect in ordinary matter. We also quantify the enhanced effect produced by the speculated \emph{Heisenberg-Couloumb} (H-C) effect in superconductors, thereby providing a test for the validity of the H-C theory, and consequently the existence of gravitons. 
\end{abstract}


\maketitle

\begin{figure}
	\centering
	\includegraphics[width=.8\columnwidth]{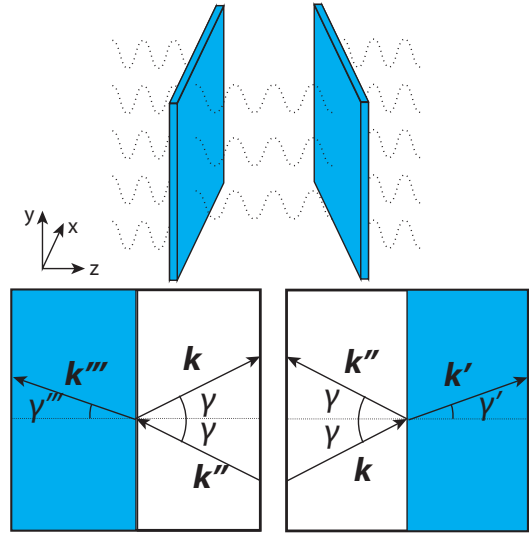}
	\caption{The Casimir effect with a two plate setup. The change in the refractive index of the plates causes the gravitational wave to refract. $\mathbf{k}$ represents the wave vector of the incident, transmitted, and reflected gravitational waves, and $\gamma$ is the corresponding angle with respect to the surface normal.}
	\label{fig:casimir}
\end{figure}

One of the most remarkable consequences of the non-zero vacuum energy predicted by quantum field theory, is the Casimir effect. In its most basic form, the Casimir effect is the attraction between two perfectly reflecting surfaces as a result of the restriction of allowed modes in the vacuum between them~(Fig.~\ref{fig:casimir}). Real bodies however are not perfectly reflecting, and the generalisation of these ideal boundary conditions to more realistic ones have been derived for the electromagnetic (EM) field, resulting in the Lifshitz formula at zero temperature~\cite{lifshitz56}. The EM field of course, is not the only field that produces the Casimir effect; in theory all fields of the quantum vacuum contribute to the Casimir effect. In fact the contribution to the Casimir effect from any massless field which is opaque to the plates should be significant (mass quickly weakens the Casimir effect~\cite{hays79,*ambjorn83}). Therefore one may imagine plates which are opaque to the gravitational field, so that the Casimir effect would then be a manifestation of the quantisation of the gravitational field, or gravitons. The difficulty is in finding such a medium, as ordinarily materials are transparent to the gravitational field~\cite{gillies87}. 

Recently however, there have been suggestions that the properties of \emph{quantum fluids} (superconductors, superfluids, quantum Hall fluids, Bose-Einstein condensates) may enhance the interaction with gravitational waves (GW). The novel effects of the interaction of a gravitational field with a quantum fluid was first investigated by DeWitt~\cite{dewitt66} and Papini~\cite{papini66}, who calculated that a Lense-Thirring field should induce a current in the superconductor. Following this, further analyses were made into the interaction of GW with superconductors~\cite{peng90,peng91}, proposing superfluids as a medium for gravitational antennae~\cite{anandan81,*chiao82,*anandan82,*anandan84}, superconducting circuits as GW detectors~\cite{anandan85}, transducers~\cite{chiao03, licht04} and mirrors~\cite{minter10}. These idea have not been met without controversy~\cite{kowitt94,*harris99,keifer05}. Although a few experiments have attempted to test the proposed enhanced GW interaction~\cite{podkletnov92,*tajmar07,*tajmar08,chiao03a,*podkletnov03}, none have produced clear and unambiguous outcomes. This is perhaps because of the small magnitude of some of the theorised effects coupled with the practical challenges in producing the environment capable of their detection~\cite{chiao03a}. In light of the controversial status of enhanced gravitational interaction, an understanding of the gravitonic Casimir effect for realistic bodies is needed to accurately quantify the contribution in ordinary materials and in any theorised system with enhanced gravitational interactions.

The Casimir effect has also been investigated in weak gravitational fields to see the effect the slightly curved spacetime background would have on the Casimir energy~\cite{calloni02,*caldwell02,*sorge05,*bimonte08a}. In these works, the gravitational field is assumed to be unaffected by the matter that form the boundary conditions. This is different from what we are considering here, where we look at how the gravitational field of the vacuum interacts with the matter of the boundary conditions to give rise to the gravitonic Casimir effect, in flat spacetime. This gravitonic Casimir effect has been considered in a cosmological context; however in these works, the boundary conditions are idealised and not suitable for realistic terrestrial systems~\cite{lin01,*ruser07}. The gravitonic Casimir potential also has been calculated for a massive test point particle interacting with a fluctuating mass distribution~\cite{panella94}. In this letter we derive the gravitonic Casimir effect for real bodies. In effect we give the Lifshitz formula for the gravitonic contribution to the Casimir effect at zero temperature.

The contribution of small perturbations to the flat spacetime metric, $\eta_{\mu\nu}$, is well described by the linearised Einstein field equations. We begin with a Maxwell-like formulation of the linearised Einstein field equations known as \emph{gravitoelectromagnetism} (GEM)~\cite{matte53,*campbell71,*maartens98,*ramos10,szekeres71}, 
\begin{gather}
\nabla\cdot\mathbf{E}=\kappa\boldsymbol{\rho}^{(E)}~,\label{gem_1}\\
\nabla\cdot\mathbf{B}=\kappa\boldsymbol{\rho}^{(M)}~,\label{gem_2}\\
\nabla\times\mathbf{E}=-\frac{\partial\mathbf{B}}{\partial t}-\kappa\mathbf{J}^{(M)}~,\label{gem_3}\\
\nabla\times\mathbf{B}=\frac{\partial\mathbf{E}}{\partial t}+\kappa\mathbf{J}^{(E)}~.\label{gem_4}
\end{gather}
where $\kappa\equiv8\pi G/c^4$. In this formulation, components of the Weyl tensor ($C_{\alpha\beta\mu\nu}$) play roles analogous to the electric and magnetic fields in electromagnetism: $E_{ij} \equiv C_{0i0j}$ and $B_{ij}\equiv {\star C_{0i0j}}$,  ( $\star$ denotes Hodge dualisation~\cite{dual}). The $rhs$ of the GEM equations  contain components of the matter current, $J_{\mu\nu\rho}\equiv-T_{\rho[\mu,\nu]}+\frac13\eta_{\rho[\mu}T_{,\nu]}$ where $T_{\mu\nu}$ ($T\equiv T^\mu_\mu$) are the stresses due to the perturbation: ${\rho^{(E)}_i\equiv-J_{i00}}$, ${\rho^{(M)}_i\equiv{-\star J_{i00}}}$, ${J^{(E)}_{ij}\equiv J_{i0j}}$, and ${J^{(M)}_{ij}\equiv\star J_{i0j}}$ (we use the convention that Greek letters go from 0 to 3 and Roman letters go from 1 to 3, unless otherwise stated). 

We are interested in the macroscopic properties of the system, so $\mathbf{E}, \mathbf{B}, \mathbf{T}$ will represent Russakoff spatial averaged values~\cite{russakoff70}, as is used in the macroscopic Maxwell's equations. We assume that $T_{ij}=\chi(\omega) E_{ij}$, where $\chi(\omega)$ is the frequency dependent \emph{gravitational susceptibility}~\cite{ingraham97};  we will later show a specific  model of matter where the induced stresses are of this form. We consider the gravitoelectromagnetic fields to be of plane wave form, $E_{ij}(\mathbf{r},t) = \mathcal{E}_{ij}e^{i(\mathbf{k}\cdot\mathbf{r}-\omega t)}$, $B_{ij}(\mathbf{r},t) = \mathcal{B}_{ij}e^{i(\mathbf{k}\cdot\mathbf{r}-\omega t)}$. From the GEM equations, $\mathbf{E}$ and $\mathbf{B}$ are transverse waves with two independent polarisations, `+' and `$\times$', respectively defined by the only non-vanishing components $\mathcal{E}_{11}=-\mathcal{E}_{22}=\mathcal{B}_{12}=\mathcal{B}_{21}= \alpha$ and $-\mathcal{B}_{11}=\mathcal{B}_{22}=\mathcal{E}_{12}=\mathcal{E}_{21}= \beta$, in the \emph{proper frame} of the plane wave. Here we define the proper frame of the plane wave as that in which $\mathbf{k}$ is along the positive $z$-axis.

The vacuum energy of any quantum field between parallel plates (separated by distance $a$) is a summation of the energy of all allowed modes of the field. The opaque boundary of the parallel plates in the $xy$-plane means that the $k_z$ components of the field are discrete, whereas the $\mathbf{k}_{\parallel}\equiv(k_x,k_y)$ components remain continuous. In terms of the graviton eigenfrequencies, the vacuum energy of gravitons between the plates at zero temperature is given by~\cite{bordag09},  
\begin{equation}
E_0=\frac{\hbar}{4\pi}\int_0^\infty k_{\parallel} dk_{\parallel} \sum_{n}(\omega_n^++\omega_n^\times)\sigma~\label{E0},
\end{equation}
where $\sigma$ is the surface area of the plate. The allowed modes ($\omega_n^+$, $\omega_n^\times$) between the plates are found by considering the boundary conditions, as follows.

A naive application of Stokes' theorem to the $\mathbf{E}$ and $\mathbf{B}$ fields at the plate interface will produce an overdetermined problem. The reason for this is that the extra components of these second-order tensors would introduce extra constraints, as compared to vector fields, such as the EM field. Thus $\mathbf{E}$ and $\mathbf{B}$ cannot be considered completely smooth across the interface. Instead only the traceless part of the tangential components of the tensor fields are considered smooth across the interface - this is known as the \emph{smoothness principle}~\cite{ingraham97}. This gives rise to the  boundary conditions, 
\begin{gather}
\Delta[(1+\kappa\chi/2)\mathbf{E}^{\text{TT}}]=0~,\label{E_bc}\\
\Delta \mathbf{B}^{\text{TT}}=0\label{B_bc}~,
\end{gather}
where $E_{ab}^{TT} = \mathcal{E}_{ab}^{TT}e^{i(\mathbf{k}\cdot\mathbf{r}-\omega t)}$ is the traceless part of $\mathbf{E}$ after it has been projected onto the interface, and $B_{ab}^{TT}= \mathcal{B}_{ab}^{TT}e^{i(\mathbf{k}\cdot\mathbf{r}-\omega t)}$ is the traceless part of $\mathbf{B}$ after it has been projected onto the interface (subscripts $a,b=1,2$). In the interface frame as shown in Fig.~(\ref{fig:casimir}),   $\mathcal{E}_{11}^{TT}=-\mathcal{E}_{22}^{TT}=\alpha(1-S^2/2)$, $\mathcal{E}_{12}^{TT}=\mathcal{E}_{21}^{TT}=\beta C$, and $\mathcal{B}_{11}^{TT}=-\mathcal{B}_{22}^{TT}=-\beta(1-S^2/2)$, $\mathcal{B}_{12}^{TT}=\mathcal{B}_{21}^{TT}=\alpha C$, where $S\equiv\sin\gamma=k_{\parallel}/k$ and $C\equiv\cos\gamma=k_z/k$.  $\Delta Q \equiv Q_1 - Q_2$ refers to the change in quantity $Q$ at the interface between medium 1 and medium 2.

We now adapt van Kampen \emph{et al.}'s~\cite{vanKampden68} contour integral method, except with different boundary conditions, to get the gravitonic Casimir energy. Applying the boundary conditions of Eq.~(\ref{E_bc}) and (\ref{B_bc}) at the two plate interfaces, we get the following system of linear homogeneous equations of variables $\alpha,\alpha',\alpha'',\alpha'''$ for the `+'-polarisation (primes indicate the region of operation, as represented in Fig.~\ref{fig:casimir}),
\begin{widetext}
\begin{align}
\alpha'(1+\kappa\chi/2)(1-S'^2/2)e^{-q'a/2}&=\alpha(1-S^2/2)e^{-qa/2}+\alpha''(1-S^2/2)e^{qa/2}~,\label{soe_1}\\
\alpha'''(1+\kappa\chi/2)(1-S'^2/2)e^{-q'a/2}&=\alpha(1-S^2/2)e^{qa/2}+\alpha''(1-S^2/2)e^{-qa/2}~,\label{soe_2}\\
\alpha'C'e^{-q'a/2}&=\alpha Ce^{-qa/2}-\alpha''Ce^{qa/2}~,\label{soe_3}\\
-\alpha'''C'e^{-q'a/2}&=\alpha Ce^{qa/2}-\alpha''Ce^{-qa/2}~,\label{soe_4}
\end{align}
\end{widetext}
where $q^2\equiv-k_z^2=k_{\parallel}^2-\omega^2/c^2$, and  $q'^2\equiv-k_z'^2=k_{\parallel}^2-(1+\kappa\chi)\omega^2/c^2$. 
A non-trivial solution exists when the determinant of the corresponding matrix is zero,
\begin{multline}
\Delta^+\equiv e^{-aq'}\{[C'(S^2-2)-(1+\kappa\chi/2)C(S'^2-2)]^2e^{-aq}\\
-[C'(S^2-2)+(1+\kappa\chi/2)C(S'^2-2)]^2e^{aq}\}=0~.\label{Delta_+}
\end{multline}
The boundary conditions also yield a similar set of linear homogeneous equations of variables $\beta,\beta',\beta'',\beta'''$, from which we get for the `$\times$'-polarisation, 
\begin{multline}
\Delta^\times\equiv e^{-aq'}\{[(1+\kappa\chi/2)C'(S^2-2)-C(S'^2-2)]^2e^{-qa}\\
-[(1+\kappa\chi/2)C'(S^2-2)+C(S'^2-2)]^2e^{qa}\}=0~.\label{Delta_x}
\end{multline}

Solutions of Eq.~(\ref{Delta_+}) and (\ref{Delta_x}) give the allowed modes between the plates, i.e. the eigenfrequencies ($\omega_n^+$, $\omega_n^\times$). There will be an infinite number of these eigenfrequencies. We can sum them in Eq.~(\ref{E0}), using the \emph{argument principle} of complex analysis~\cite{ahlfors79},

\begin{equation}
\sum_n\omega_n=\frac{1}{2\pi i}\left[\int\limits_{i\infty}^{-i\infty}\omega d\ln\Delta+\int\limits_{C^+}\omega d\ln\Delta\right]\label{sum_omega}~,
\end{equation}
where the closed contour of integration has been taken over a semicircle $C^+$ of infinite radius in the right half of the complex plane of $\omega$ with its centre at the origin, and the imaginary axis $[i\infty,-i\infty]$, in a counterclockwise manner.

In the high frequency limit, the system does not have time to respond to the rapid oscillations of the field, and therefore will effectively act with time averaged behaviour. Analogous to the EM case, we thus make the natural assumption $\lim_{w\to\infty}\chi(\omega)=\lim_{w\to\infty}d\chi(\omega)/d\omega=0$ and therefore $\gamma=\gamma'$ in this limit i.e. at very high frequencies the plates are effectively transparent to the GW. With this assumption, the second integral in \eq{sum_omega} is independent of $a$.

The summation over the infinite number of allowed modes will of course give an infinite value for $E_0$. To get the finite Casimir energy per unit area we take away the energy at infinite separation. Employing the argument principle to Eq.~(\ref{E0}) and then integrating by parts, one arrives at the gravitonic Casimir energy ($\xi \equiv -i\omega$),
\begin{equation}
\begin{split}
E(a)&=\frac{E_0}{\sigma}-\lim_{a\to\infty}\frac{E_0}{\sigma}\\
&=\frac{\hbar}{4\pi^2}\int_0^\infty k_\parallel dk_\parallel\int_0^\infty[\ln(1-r_+^2e^{-2qa})\\
&\quad\quad\quad\quad\quad\quad\quad\quad+\ln(1-r_\times^2e^{-2qa})]d\xi~,\label{E_1}
\end{split}
\end{equation}
where,
\begin{align}
r_+&\equiv\frac{C'(S^2-2)-(1+\kappa\chi/2)C(S'^2-2)}{C'(S^2-2)+(1+\kappa\chi/2)C(S'^2-2)}~,\label{r_+}\\
r_\times&\equiv \frac{(1+\kappa\chi/2)C'(S^2-2)-C(S'^2-2)}{(1+\kappa\chi/2)C'(S^2-2)+C(S'^2-2)}~.\label{r_x}
\end{align}

The boundary conditions Eq.~(\ref{E_bc}) and (\ref{B_bc}) at an interface yield Eq.~(\ref{soe_1}) and (\ref{soe_3}). Simultaneously solving Eq.~(\ref{soe_1}) and (\ref{soe_3}) one gets 
\begin{equation}
\frac{\alpha''}{\alpha}=\frac{-C'(S^2-2)+(1+\kappa\chi/2)C(S'^2-2)}{C'(S^2-2)+(1+\kappa\chi/2)C(S'^2-2)}e^{qa}\label{alpha_ratio}~.
\end{equation}
Comparison of Eq.~(\ref{alpha_ratio}) with Eq.~(\ref{r_+}) shows that, $|r_+|=|\alpha''/\alpha|$, i.e. $r_+$ is the magnitude of the gravitational reflection coefficient of the `$+$'-polarisation. Similarly for the `$\times$'-polarisation, $|r_\times|=|\beta''/\beta|$ is the magnitude of the gravitational reflection coefficient of the `$\times$'-polarisation. This tells us that Eq.~(\ref{r_+}) and (\ref{r_x}) are the Fresnel reflection coefficients for the two polarisations of the gravitational wave. Therefore Eq.~(\ref{E_1}) has the same form as the Lifshitz formula for the EM Casimir energy at zero temperature, except the EM reflection coefficients have been replaced with their gravitational equivalent. It is important to point out that this was not a priori obvious, as our starting point was the linearised Einstein equations, which is fundamentally different from Maxwell's equation of electromagnetism, although GEM provides useful analogies between the two theories. A marked difference is that the EM fields are vectors whereas $\mathbf{E}$ and $\mathbf{B}$ are tensor fields.

We need now to find an estimate for $\chi(\omega)$. A number of models of the interaction of GW with bulk matter exists. For example, Ref.~\cite{szekeres71,grishchuk80} considered a medium consisting of molecules modelled as individual harmonic oscillators to calculate the quadrupole moment induced by an incident GW; Ref.~\cite{polnarev72,*chesters73} studied the interaction and dispersion of GW in a hot gas. Here we use Peters'~\cite{peters74} model who considered the scattering of GW by the gravitational field of individual free particles of a thin sheet of matter. Peters derive a gravitational refractive index $n$ which was much larger than that generated by just considering the induced quadrupole moments, suggesting that his model encapsulates the dominant GW interaction with matter. Peters gives the gravitational refractive index as,
\begin{equation}
	n=1+\frac{2\pi G\rho}{\omega^2}~,
\label{n}
\end{equation}
where $\rho$ is the density of the medium.

Under the Lorentz gauge (${h^{\mu\alpha}}_{,\alpha}=0$), $\partial^\alpha\partial_\alpha\ h_{\mu\nu}=-2\kappa T_{\mu\nu}$ (note $h_{\mu\nu}$ is traceless). Using this, Eq.~(\ref{n}), and the relation $E_{ij}=-\omega^2 h_{ij}/2$~, result in gravitational stresses of the previously assumed form, $T_{ij}=\chi(\omega)E_{ij}$ where,
\begin{equation}
	\chi(\omega) = \frac{1-n(\omega)^2}{\kappa c^2}~,
\end{equation}
with the high frequency limits: $\lim_{w\to\infty}\chi(\omega)=\lim_{w\to\infty}d\chi(\omega)/d\omega=0$.

Except for the lowest frequencies, for ordinary material parameters, $\chi\ll c^4/G$, and therefore the reflection coefficients and Casimir pressure [$P(a)=-{\partial E(a)}/{\partial a}$] are negligible. Specifically, if we take a typical $O[\rho] = 10^4$kg/m$^3$ and $O[l]=1$\AA, then  $P(10^{-6})\approx10^{-21}$nPa. This value is of course beyond detection. For the Casimir pressure to be measurable, the density of the material would have to be at the very least $O[\rho]=O[c^2/G]\approx10^{27}$kg/m$^3$. This material density is clearly not achievable, at least terrestrially. 

Up until now we have only considered materials with classical properties. Recently,  Minter \emph{et al.}~\cite{minter10} propose that the quantum mechanical properties of superconducting films can give rise to specular reflection of GW. Their claim is that in an ordinary metal plate, the ions and \emph{normal} electrons locally co-move together along the same geodesics in the presence of a GW. However, when the plate becomes superconducting, the quantum-mechanical non-localisability of the negatively charged Cooper pair undergoes non-geodesic motion, whereas the positive charged ions of the lattice remains on the geodesic path. Specifically, the authors couple the gravitational field to the superconductor through DeWitt's minimal coupling scheme~\cite{dewitt66}. To first-order, the ground state of the delocalised Cooper pairs do not change in the presence of a GW whose frequency is less than the BCS gap frequency. In comparison, the shift in the momentum of the localised ions is proportional to the gravitational vector potential. Because the Cooper pairs and ions are oppositely charged, a strong Coulomb force will resist this separation of charge caused by the GW, resulting in its reflection. The authors dub this the \emph{Heisenberg-Coulomb} effect. A similar effect had been proposed in Ref.~\cite{peng90}.

The origins of the arguments employed by Minter \emph{et al.} are heuristical in nature, some of which we believe require a much more formal approach to be convincing. This is echoed in a review article~\cite{keifer05} on theories of enhanced gravitational interaction with quantum fluids, which predates Ref.~\cite{minter10}; in particular the authors urge that a formal derivation of the quantum mechanical coupling between an electron and both the electromagnetic and gravitational field is needed. Nevertheless, the work by Minter \emph{et al.} do yield results which can be used to falsify their theory. The H-C effect should enhance the Casimir pressure between superconducting plates. Here we quantify the size of this effect.

Minter \emph{et al.} give the reflection coefficient of a superconducting film from an incident GW as, 
\begin{equation}
r_G=\left(1+\frac{2\delta^2}{cd}\xi\right)^{-1}~,\label{rG}
\end{equation}
where $\delta$ is the EM skin depth of the superconducting film and $d$ the film thickness. In the thin superconducting film, the current flows in the $x-y$ plane; therefore only the normal incident component of the GW will drive the current. At normal incidence the magnitude of the reflection coefficient of the two polarisation modes are the same, as they only differ by a rotation (this is also true for EM waves). We can use Eq.~(\ref{rG}) in Eq.~(\ref{E_1}) to calculate the gravitonic Casimir pressure between two superconducting films.
	
We calculate the gravitonic contribution to Casimir pressure for superconducting lead (Pb) of thickness $d=2$ nm at zero temperature. The EM skin depth of Pb is $\delta=37$ nm. We compare this with the photonic contribution to the Casimir pressure of superconducting lead. The EM reflection coefficient is~\cite{minter10}
\begin{equation}
r_E=\left(1+\frac{2\lambda\delta^2}{cd^2}\xi\right)^{-1}~,\label{rE}
\end{equation}
where $\lambda=83$ nm is the coherence length. The photonic contribution to the Casimir pressure is calculated by using Eq.~(\ref{rE}) in the EM Lifshitz formula~\cite{lifshitz56}, which has the same form as Eq.~(\ref{E_1}).

\begin{figure}
	\centering
	\includegraphics[width=0.8\columnwidth]{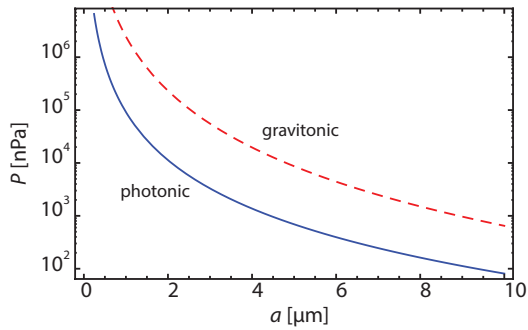}
	\caption{Gravitonic (dotted line) and photonic (solid line) contributions to the Casimir pressure of parallel plates of superconducting Pb at zero temperature, as a function of plate separation $a$. Due to the Heisenberg-Coulomb effect, the gravitonic contribution exceeds the photonic.}
	\label{pressure}
\end{figure}

Eq.~(\ref{rE}) and (\ref{rG}) are most valid when the driving frequency is less than the BCS gap frequency, as for higher frequencies the dissipative component of the complex conductivities would need to be taken into account. However, as higher frequency modes’ contribute exponentially less to the Casimir energy, one may still use the simple reflection coefficients derived by Minter \emph{et al.} to obtain the first-order estimate of the Casimir pressure for superconductors at zero temperature. \fig{pressure} compares the gravitonic to photonic contribution of the Casimir pressure as a function of separation of plates of superconducting Pb. It shows that gravitons can have a significant contribution to the Casimir pressure, via the H-C effect. In fact, for the superconducting Pb film considered here, the gravitons will dominate the Casimir pressure by an order of magnitude over photons.

The magnitude of the Casimir pressure and plate separation distance that we are talking about here, is comparable in size to what has already been achieved in current  experiments~\cite{harris00,*decca07,bressi02}. Few experiments however, have been conducted at low temperatures~\cite{chiu08,*decca10,*laurent12,*castillo13}, as room temperature setups are a more experimentally accessible environment. Of these low temperature investigations, only the ALADIN project has experimented with superconducting aluminium (Al) film, separated by a thin oxide layer from a thick gold plate, to observe how the Casimir energy influences the superconducting phase transition (preliminary experimental results are reported in~\cite{bimonte08,*alloca12}). One could imagine modifying such an experiment to test for the gravitonic contribution to the Casimir effect as described in this letter. 

We have derived here a Lifshitz-type formula for the gravitonic Casimir effect for real bodies. Besides completing a theoretical gap in our understanding of the Casimir effect, this formula is important in light of recent models of enhanced gravitational interaction, as it allows us to quantify the gravitonic contribution to the Casimir effect predicted by these theories. If measurements of the Casimir pressure of the setup described in this letter match only to a photonic contribution [Fig.~(\ref{pressure}) solid line], then one should conclude that the H-C effect is invalid, if we are to hold on to the idea of the graviton. However, if experiments show the Casimir pressure to be an order of magnitude larger than that predicted from the photonic contribution alone, this would be the first experimental evidence for the validity of the H-C theory and the existence of gravitons. This would open a new field in the way of graviton detection.

This work was financially supported by the Japan Society for the Promotion of Science. The author is an International Research Fellow of the Japan Society for the Promotion of Science.

\bibliography{casimir_prl}

\begin{thebibliography}{55}
\bibitem [{dua()}]{dual}%
  \BibitemOpen
  \href@noop {} {}\bibinfo {note} {The folllowing dualisation is defined:
  $\star
  C_{\mu\nu\rho\sigma}\equiv\frac12\epsilon_{\mu\nu\alpha\beta}C^{\alpha\beta}_{\rho\sigma}$
  and $\star
  J_{\mu\nu\rho}\equiv\frac12\epsilon_{\mu\nu\alpha\beta}J^{\alpha\beta}_{\rho}$,
  where $\epsilon_{\mu\nu\alpha\beta}$ is the Levi-Civita tensor.}\BibitemShut
  {Stop}%
\end{thebibliography}

\end{document}